\documentclass{aa}
\usepackage{graphicx}
\usepackage{aas_macros}
\usepackage{txfonts}
\usepackage{natbib}
\def\sw {\emph{Swift}}
\def\swx {\emph{Swift}/XRT}
\def\src {LS\,I\,+61$^\circ$\,303}
\def\ls {LS\,5039}
\def\psr {PSR\,B1259$-$63}
\def\azero {A\,0538$-$66}
\def\flux {\mbox{erg cm$^{-2}$ s$^{-1}$}}
\def\lum {\mbox{erg s$^{-1}$}}

\begin{document}
   \title{\swx\ monitoring of five orbital cycles of \src}

   \author{P. Esposito\inst{1,2}
   \and P. A. Caraveo\inst{2} 
   \and A. Pellizzoni\inst{2}
   \and A. De Luca\inst{2}  
   \and N. Gehrels\inst{3}
   \and M.~A.~Marelli\inst{2}
}

   \offprints{Paolo Esposito, paoloesp@iasf-milano.inaf.it}
   \institute{Universit\`a degli Studi di Pavia, Dipartimento di Fisica Nucleare e Teorica and INFN-Pavia, via Bassi 6, I-27100 Pavia, Italy
    \and INAF - Istituto di Astrofisica Spaziale e Fisica Cosmica Milano, via Bassini 15, I-20133 Milano, Italy
    \and NASA/Goddard Space Flight Center, Greenbelt, Maryland 20771, USA
}
   \date{Received / Accepted}

\abstract
{\src\ is one of the most interesting high-mass X-ray binaries owing to its spatially resolved radio emission and its TeV emission, generally attributed to non-thermal particles in an accretion-powered relativistic jet or in the termination shock of the relativistic wind of a young pulsar. Also, the nature of the compact object is still debated. Only \ls\ and \psr\ (which hosts a non-accreting millisecond pulsar) have similar characteristics.}
{We study the X-ray emission from \src\ covering both short-term and orbital variability. We also investigate the source spectral properties in the soft X-ray (0.3--10 keV) energy range.}
{Twenty-five snapshot observations of \src\ were collected in 2006 with  the XRT instrument on-board the \sw\ satellite over a period of four months, corresponding to about five orbital cycles. Since individual data sets have too few counts for a meaningful spectral analysis, we extracted a cumulative spectrum.}
{The count rate folded at the orbital phase shows a clear modulation pattern at the 26.5 days period and suggests that the X-ray peak occurs around phase 0.65. Moreover, the X-ray emission appears to be variable on a timescale of $\sim$1 ks. The cumulative spectrum is well described by an absorbed power-law model, with hydrogen column density $N_{\rm H}=(5.7\pm0.3)\times10^{21}$ cm$^{-2}$ and photon index $\Gamma=1.78\pm0.05$. No accretion disk signatures, such as an iron line, are found in the spectrum.}
{}
\authorrunning {P. Esposito et al.}
\titlerunning {\src\ observed by Swift}
\maketitle
\section{Introduction}
\src\ is a peculiar binary system characterized by variable radio emission, long known to be modulated with a \mbox{$26.4960 \pm 0.0028$ days} orbital period \citep{gregory78,gregory02}. The eccentricity of the orbit is $0.72\pm0.15$, and the periastron passage occurs at phase $\phi=0.23\pm0.02$ \citep{casares05}. The phase of the radio maximum shifts between $0.45 \leq \phi\leq 0.9$ with a \mbox{$1667\pm8$ days} period \citep{gregory02}, pointing to a second modulation. \ion{H}{i} observations of the radio source give a distance of \mbox{$2.0 \pm 0.2$ kpc} \citep{frail91}, implying an X-ray luminosity of \mbox{$\sim$$10^{33}$ \lum}. Optical studies have established the primary star to be a Be surrounded by a small and inhomogeneous disk whose optical and \ion{H}{$\alpha$} emission is also modulated \citep{hutchings81,mendelson89,paredes94,zamanov99}.\\
\indent \src\ has been attracting a lot of attention (and observing time) owing to its putative link with the gamma-ray source 2CG\,135+01, discovered by  \emph{COS-B} \citep{hermsen77,swanenburg81,bignami83} 30 years ago and later confirmed by  \emph{CGRO}/EGRET, as 3EG\,J0241+6103 \citep{hartman99}, with observations spanning the period 1991--2000. In spite of relentless efforts \citep{tavani96,kniffen97}, no conclusive evidence for the orbital modulation was found in the gamma-ray data, thus hampering the source identification. Recently the orbital modulation was seen by the  \emph{MAGIC} Cherenkov air-showers telescope in ultra high energy gamma rays (above 400 GeV), clinching the case for the identification of \src\ as a gamma-ray emitter \citep{albert06}. This was also confirmed in the hard X-ray energies (up to \mbox{100 keV}) by \emph{INTEGRAL} observation of variable emission, peaking at phase $\sim$0.55 of the orbital modulation \citep{chernyakova06,hermsen07}. \\
\indent Although in soft X-rays there were no doubts from the start about the source identification \citep{bignami81}, the X-ray
observations, aimed at studying the orbital modulation, proceeded in parallel with the gamma-ray ones, using all the X-ray
instruments active over the past 30 years: \emph{Einstein} \citep{bignami81}, \emph{ROSAT} \citep{goldoni95,taylor96}, \emph{ASCA} \citep{leahy97}, \emph{RossiXTE} \citep{paredes97,harrison00,greiner01,leahy01,grundstrom07}, \emph{INTEGRAL} \citep{chernyakova06,hermsen07}, \emph{BeppoSAX}, \emph{XMM-Newton} \citep{sidoli06}, and \emph{Chandra} \citep{paredes07}.\\
\indent Today, the behavior of \src\ continues to defy easy characterization. Its X-ray emission appears to be variable over both long and short time scales \citep[e.g.,][]{sidoli06}, hampering the detection of an unambiguous modulation pattern when only sparse observations are available. Here we report the monitoring of the source soft X-ray flux over 5 adjacent orbital cycles, using the XRT telescope onboard the NASA/UK/ASI \sw\ mission \citep{gehrels04}.
%
%
\section{Observations and analysis}\label{data}
\begin{figure}[h!]
\resizebox{\hsize}{!}{\includegraphics[angle=-90]{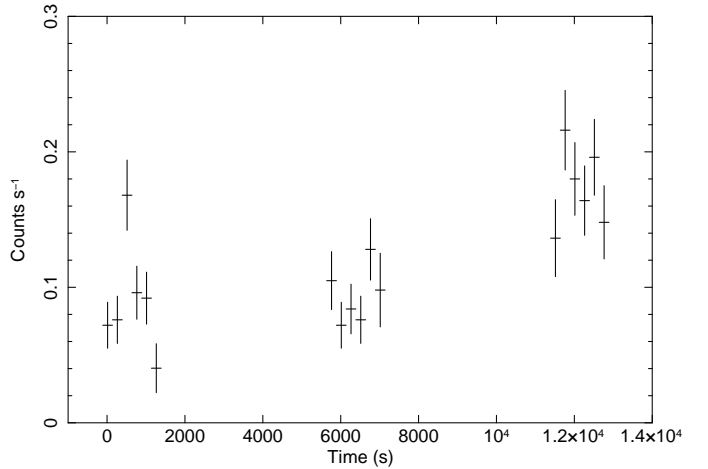}}
\caption{\swx\ light curve of the first 14 ks of observation 00036093003 (see Table \ref{journal} for details). The bin size is 250 s and less than \mbox{0.001 counts s$^{-1}$} are expected from background events. The short-term X-ray variability of \src\ is apparent.}
\label{obs3}
\end{figure}

\begin{figure}[h!]
\resizebox{\hsize}{!}{\includegraphics[angle=-90]{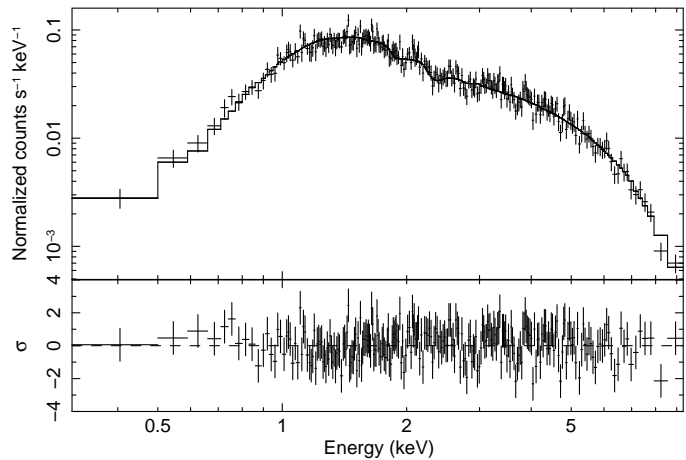}}
\caption{\swx\ cumulative spectrum of \src\ (see Sect.\,\ref{data} for details). \emph{Top:} data and best-fit power-law model. \emph{Bottom:} residuals from the best-fit model in units of standard deviation.}
\label{spec}
\end{figure}

\begin{figure}[h!]
\resizebox{\hsize}{!}{\includegraphics[angle=0]{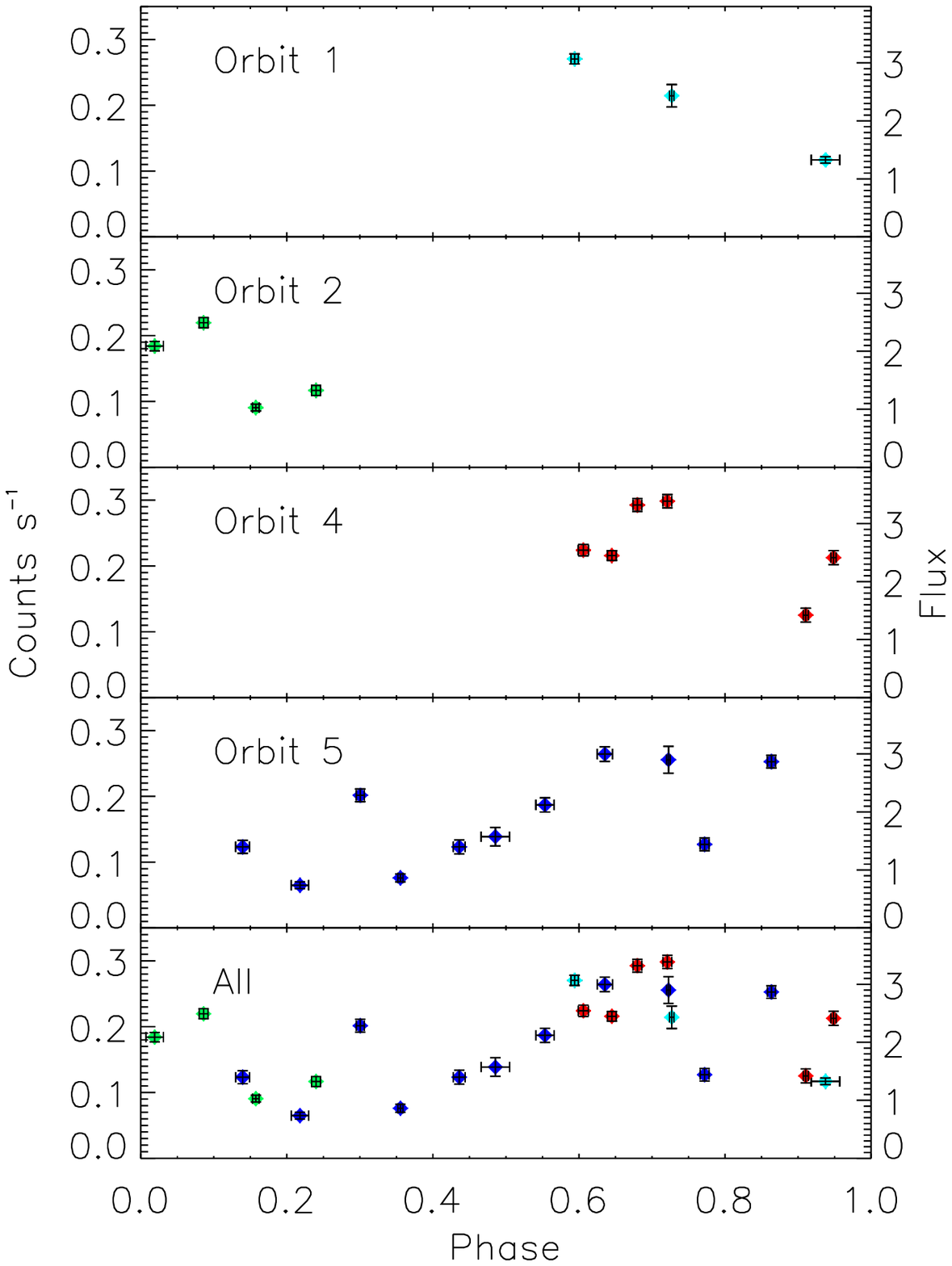}}

\vspace{-.5cm}

\resizebox{\hsize}{!}{\includegraphics[angle=0]{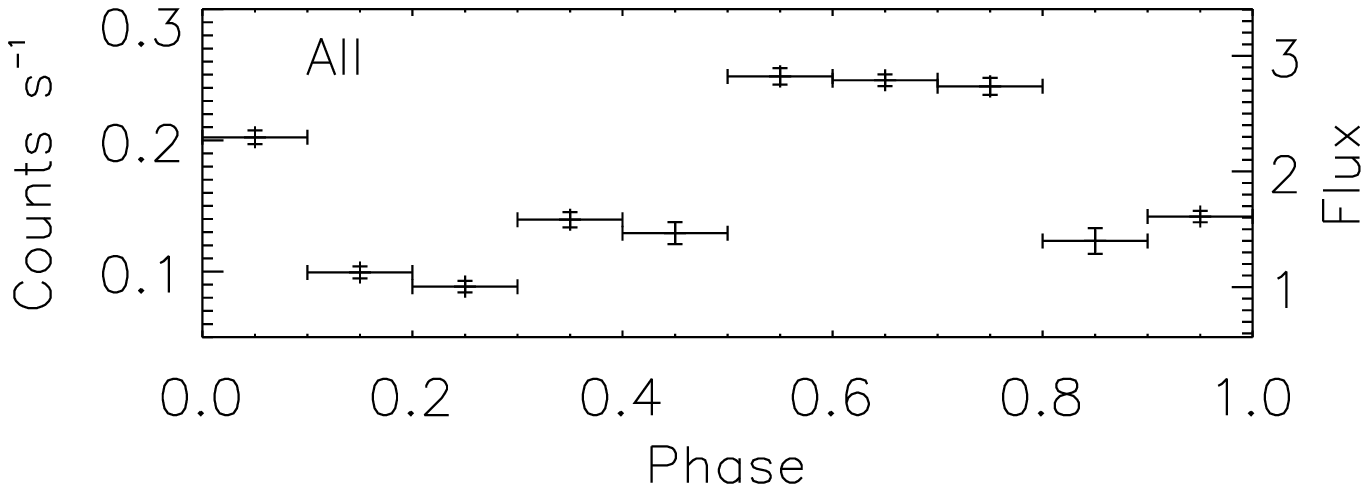}}
\caption{\swx\ background-subtracted count rates folded as a function of the radio orbital phase (in the 0.3--10 keV energy range). The count rates have been converted to unabsorbed fluxes (in units of \mbox{$10^{-11}$ \flux}) in the same energy range using the conversion factor given in Section \ref{data}. \emph{Top:} on each panel, the data, averaged over a single observation, are plotted for different orbits (in the fifth panel all data are plotted simultaneously). \emph{Bottom:} all data binned at 0.1 in phase are plotted simultaneously. See the online edition of the article for a color version of this figure.}
\label{countslc}
\end{figure}

Twenty-five \sw\ observations of \src\ have been performed with the XRT instrument in photon counting (PC) mode \citep{burrows05}. The XRT uses a CCD detector sensitive to photons with energies between 0.2 and 10 keV. Table \ref{journal} reports the log of the \swx\ observations used for this work.
 \begin{table*}
 \begin{center}
 \caption{Observation log.}
 \label{journal}
 \begin{tabular}{cccccc}
 \hline
 \hline
 \noalign{\smallskip}
Sequence & Start time (UT) & End time (UT) & Phase$^{\mathrm{a}}$ & Net exposure$^{\mathrm{b}}$ & Count rate$^{\mathrm{c}}$ \\
 & (yyyy-mm-dd hh:mm:ss) & (yyyy-mm-dd hh:mm:ss)& & (s) & (counts s$^{-1}$) \\
 \noalign{\smallskip}
 \hline
 \noalign{\smallskip}
00036093001 & 2006-09-02 08:09:31 & 2006-09-02 11:46:56 & 0.591--0.597 & 4665 & $0.271\pm0.008$ \\
00036093002 & 2006-09-05 21:39:40 & 2006-09-05 23:22:56 & 0.726--0.728 &   749 & $0.22\pm0.02$ \\
00036093003 & 2006-09-11 00:57:04 & 2006-09-11 23:30:56 & 0.920--0.955 &   4657 & $0.120\pm0.005$ \\
00036093004 & 2006-09-13 09:40:21 & 2006-09-13 22:36:57 & 0.009--0.029 &   3598 & $0.187\pm0.007$ \\
00036093005 & 2006-09-15 07:56:34 & 2006-09-15 13:11:57 & 0.082--0.090 &   3783 & $0.223\pm0.008$ \\
00036093006 & 2006-09-17 06:33:09 & 2006-09-17 09:54:58 & 0.155--0.160 &   3559 & $0.091\pm0.005$ \\
00036093007 & 2006-09-19 09:53:05 & 2006-09-19 14:57:57 & 0.236--0.244 &   2053 & $0.117\pm0.008$ \\
00036093008 & 2006-11-21 01:56:37 & 2006-11-21 08:22:56 & 0.601--0.611 &   3024 & $0.226\pm0.009$ \\
00036093009 & 2006-11-22 03:28:09 & 2006-11-22 08:27:56 & 0.641--0.649 &   3965 & $0.219\pm0.007$ \\
00036093010 & 2006-11-23 01:59:58 & 2006-11-23 07:01:56 & 0.676--0.684 &   3092 & $0.30\pm0.01$ \\
00036093011 & 2006-11-24 03:47:00 & 2006-11-24 08:46:56 & 0.717--0.725 &   2976 & $0.30\pm0.01$ \\
00036093012 & 2006-11-29 06:13:28 & 2006-11-29 08:02:56 & 0.909--0.912 &   1126 & $0.13\pm0.01$ \\
00036093013 & 2006-11-30 06:18:01 & 2006-11-30 08:06:56 & 0.947--0.950 &   1900 & $0.22\pm0.01$ \\
$^{\phantom{1}}$00036093014$^{\mathrm{d}}$ & 2006-12-02 22:45:00 & 2006-12-02 22:45:57 & 0.049 &   58 & $0.24\pm0.07$ \\
00036093016 & 2006-12-05 03:42:40 & 2006-12-05 13:23:58 & 0.132--0.147 &   1284 & $0.12\pm0.01$ \\
00036093017 & 2006-12-07 03:47:46 & 2006-12-07 16:51:57 & 0.208--0.228 &   2455 & $0.065\pm0.005$ \\
00036093018 & 2006-12-09 12:24:07 & 2006-12-09 16:59:57 & 0.297--0.304 &   2145 & $0.20\pm0.01$ \\
00036093019 & 2006-12-11 00:51:08 & 2006-12-11 02:48:58 & 0.354--0.357 &   2097 & $0.076\pm0.006$ \\
00036093020 & 2006-12-13 01:01:26 & 2006-12-13 09:07:57 & 0.430--0.442 &   1126 & $0.12\pm0.01$ \\
00036093021 & 2006-12-14 01:09:47 & 2006-12-14 23:40:57 & 0.468--0.503 &   707 & $0.14\pm0.01$ \\
00036093022 & 2006-12-16 01:21:46 & 2006-12-16 14:14:56 & 0.543--0.564 &   1633 & $0.19\pm0.01$ \\
00036093023 & 2006-12-18 06:15:42 & 2006-12-18 17:36:58 & 0.627--0.644 &   2168 & $0.26\pm0.01$ \\
00036093024 & 2006-12-20 19:19:35 & 2006-12-20 19:39:56 & 0.722--0.723 &   622 & $0.26\pm0.02$ \\
00036093025 & 2006-12-22 00:12:58 & 2006-12-22 05:07:58 & 0.768--0.776 &   1393 & $0.13\pm0.01$ \\
00036093026 & 2006-12-24 10:27:07 & 2006-12-24 15:21:56 & 0.860--0.867 &   2764 & $0.25\pm0.01$ \\
 \noalign{\smallskip}
  \hline
  \end{tabular}
  \end{center}
  \begin{list}{}{} 
    \item[$^{\mathrm{a}}$]  Following \citet{gregory02} we adopt as time of the zero phase JD$_{0} = 2,443,366.775$, with the orbital period $P_{\rm{orb}} = 26.4960 \pm 0.0028$ days. The mean error on the orbital phase is $\pm$0.002. The phase of the super-orbital modulation of the orbital phase and amplitude of the radio outbursts (with a period of $1667\pm8$ days; \citealt{gregory02}) spans from $\sim$0.367 (observation 00036093001) to $\sim$0.435 (observation 00036093026).
  \item[$^{\mathrm{b}}$] The exposure time is spread over several snapshots (single continuous pointings at the target) during each observation.
  \item[$^{\mathrm{c}}$] Hereafter all errors are at 1\,$\sigma$ confidence level, unless otherwise specified. 
    \item[$^{\mathrm{d}}$] This observation is listed for completeness, but will not be used for the analysis owing to its short integration time.
  \end{list}   
\end{table*}\\
\indent The data were processed with standard procedures using the FTOOLS\footnote{See \texttt{http://heasarc.gsfc.nasa.gov/docs/software/}\,.} task \texttt{xrtpipeline} (version 0.10.6 under the \texttt{Heasoft} package 6.2.0). We selected events with grades \mbox{0--12} and limited the analysis between 0.3--10 keV,  where the PC response matrices are well calibrated. We extracted the source events  from a circular region with a radius of 20 pixels \mbox{(1 pixel $\simeq 2.37''$)}, whereas to account for the background we extracted events within an annular source-free region centered on \src\ and with radii 55 and 75 pixels. Since the data show a maximum count rate $<$0.3 counts s$^{-1}$, no pile-up correction was necessary. \\
\indent Inspection of the XRT light curves relative to each observation shows evidence of moderate variability (up to a factor of $\sim$4) on a timescale of $\sim$1 ks (see Fig.\,\ref{obs3}), an effect already observed in X-rays with various instruments \citep[e.g.,][]{sidoli06}.
 However, the statistics are too poor to  assess spectral variation as a function of the source count rate.\\
\indent Individual data sets have too few counts for a meaningful spectral analysis. Therefore, we extracted a cumulative spectrum. This corresponds to a total exposure of 52.6 ks and contains about 10,000 net counts in the 0.3--10 keV range. The data were rebinned with a minimum of 20 counts per energy bin to allow $\chi^2$ fitting. The ancillary response file was generated with \texttt{xrtmkarf}, and it accounts for different extraction regions, vignetting and point-spread function corrections. We used the latest available spectral redistribution matrix (\texttt{swxpc0to12\_20010101v008.rmf}).\\
\indent The spectral fitting was performed using \texttt{XSPEC}\footnote{See \texttt{http://heasarc.gsfc.nasa.gov/docs/xanadu/xspec/}\,.} version 12.3, adopting an absorbed power-law model. We find the following best-fit ($\chi^2_{\rm{red}}=1.00$ for 359 degrees of freedom) parameters\footnote{Spectral errors are given at the 90\% confidence level for a single interesting parameter.}: absorption $N_{\rm H}=(5.7\pm0.3)\times10^{21}$ cm$^{-2}$ and photon index \mbox{$\Gamma=1.78\pm0.05$} (see Fig.\,\ref{spec}). No evidence for emission or absorption lines was found by inspecting the residuals from the best-fit models (see Fig.\,\ref{spec}). The addition of a blackbody component is not required. We used the resulting averaged flux (corrected for the absorption) of \mbox{$\sim$$2.2\times10^{-11}$ \flux} to derive the conversion factor \mbox{1 count s$^{-1}$ $\simeq$ $1.1 \times10^{-10}$ \flux} in the 0.3--10 keV energy range. The mean count rate of each observation is reported in Table \ref{journal} and the values are plotted against the orbital radio phase in Fig.\,\ref{countslc}. First we provide an orbit-by-orbit view, then all the data are plotted together and finally they are averaged over a $\sim$2.5 day bin corresponding to $1/10$  of the radio orbital phase. The source luminosity ranges from \mbox{$\sim$$3.4\times10^{33}d^2_{2}$ \lum} to \mbox{$\sim$$1.6\times10^{34}d^2_{2}$ \lum}, where we indicate with $d_{\rm{N}}$ the distance in units of N kpc.
%
\section{Discussion}
\src\ is one of the most intriguing Be/X-ray binary systems because of its periodic radio and X-ray emission and of its strong gamma-ray emission. Two main models have been proposed to account for the peculiarities of \src\ and of the similar binary \ls\ (orbital period \mbox{$P_{\rm{orb}}=3.9$ days}; \citealt{paredes00}); see \citet{mirabel06} for a recent review. The first model assumes that the systems are powered by accretion on either a neutron star or a black-hole, and that non-thermal particles are accelerated in relativistic jets. The second supposes that \src\ and \ls\ are similar to the rotation-powered system \psr, composed by a millisecond radio pulsar orbiting around a Be star ($P_{\rm{orb}}=1237$ days and pulsar spin period of 47.7 ms; e.g., \citealt{kaspi95}). In this scenario, the non-thermal emission arises from the shock resulting from the interaction between the relativistic pulsar wind and the stellar wind originating from the companion star.
The \swx\ monitoring of \src\ over five orbital periods provides more elements to constrain the parameter space of the source.\\
\indent We observed flux fluctuations up to a factor of $\approx$3 from the mean value at intra-hour scales. An example of this variability, already observed with various X-ray instruments \citep[e.g.,][]{sidoli06}, is given in Fig.\,\ref{obs3}. This kind of variability has also been observed in \ls, and it is generally interpreted as being due to variations or anisotropies in the stellar wind \citep{bosch05}.
The folding of the XRT count rates at the $\sim$26.5 days period (Fig.\,\ref{countslc}) strongly suggests that this ``flickering'' is superimposed on an overall orbital modulation. We note that the erratic behavior of \src\ should be carefully considered when assessing the orbital modulation pattern using observations of the source collected at different times.\\
\indent Previous multi-wavelength campaigns undertaken with different instruments \citep{taylor96,harrison00} indicate that the X-ray emission peak tends to precede the radio outburst by a few days. However, the exact phase of the X-ray maximum was difficult to estimate because of the poor sampling of the orbit of \src.
Recently, folding more than 600 \emph{INTEGRAL} pointings performed discontinuously with IBIS/ISGRI from March 2003 to July 2006 for a total exposure time of $\sim$1.1 Ms, and \emph{RossiXTE}/ASM data from a similar time-span, \citet{hermsen07} found a maximum at phase $\sim$0.55 in the \mbox{1--100 keV} band (in agreement with the results found by \citet{paredes97} and \citet{grundstrom07} based on \emph{RossiXTE}/ASM data only).\\
\indent By binning the folded light curves in different ways, we estimate that a broad X-ray peak occurs around phase 0.65 (see Fig.\,\ref{countslc}), while the radio outburst is expected at phase $\approx$0.8--0.9 \citep{gpt99,gregory02}. The maximum X-ray flux is not detected at periastron \citep[$\phi\simeq0.23$;][]{casares05}, where the accretion rate should reach its maximum value \citep{marti95}.\\
\indent In the framework of the model invoking a shocked pulsar wind, X-ray flux peak and radio outburst could be in principle ascribed to the same particle population. Although significant inverse Compton losses should be at work \citep[e.g.,][]{dubus06,sidoli06}, it is worth noting that the delay of $\approx$0.2 in phase ($\sim$$5\times10^5$ s) between X-ray and radio peaks is compatible with synchrotron-dominated cooling rate of electrons with typical Lorentz factor of $\gamma\sim10^5$ assuming a magnetic field of \mbox{$\sim$0.1 G} in the post-shock distribution. \citet{dubus06} predicts higher values (a few Gauss) for the magnetic field at a stand-off distance of $R_{\rm{s}}\sim10^{11}$ cm from the pulsar, but their theoretical estimate relies on as yet poorly constrained parameters such as pulsar spin-down energy and the ratio of magnetic to kinetic energy in the pulsar wind. In this context, the fact that both X-ray and radio peaks are not detected at periastron is not surprising, considering that shock strength could strongly depend on the anisotropic geometry of the pulsar wind, and not only on the separation between the two stars.\\
\indent The data collected with the XRT detector can be described by a featureless power-law spectrum with photon index \mbox{$\Gamma\sim1.8$}. Similar photon indices were also measured with other X-ray telescopes \citep[e.g.][]{sidoli06}. This value fits reasonably well in a shock acceleration scenario where electrons and positrons cool slowly \citep{chevalier00}, while spectra of high-mass accretion-powered X-ray binaries are generally harder \citep[\mbox{$\Gamma \approx 1$}; e.g.,][]{white83}.\\
\indent We remark that the spectral characteristics of \src\ are similar to those of the Be/X-ray binary \azero\ \citep{white78} in its present low-luminosity state. This system, situated in the Large Magellanic Cloud, was discovered as a source of X-ray flares showing a recurrence of 16.65 days, interpreted as the orbital period \citep{skinner80}. Optical observations also unveiled a super-orbital modulation of $\sim$421 days \citep{alcock01}. \azero\ contains a \mbox{69 ms} X-ray pulsar \citep{skinner82} and it showed iron line-emission in at least one low-amplitude flare \citep{corbet97}.\\
\indent In recent years \azero\ entered a quiescent phase, with outbursts 2--3 orders of magnitude below the early levels and low persistent luminosity \citep{campana02}. Using an \emph{XMM-Newton} observation performed in 2002, \citet{kretschmar04} found a spectrum well fitted by a featureless power-law model with $\Gamma=1.9\pm0.3$ and estimated a \mbox{0.3--10 keV} luminosity of \mbox{$(5-8)\times10^{33}$ \lum}. The mechanism presently powering the faint X-ray emission of \azero\ is unclear. It could be either a low-rate accretion or the pulsar rotational energy through the shock between the pulsar wind and the inflowing material, as proposed for \src. To our knowledge, no data concerning the radio or GeV--TeV emission of \azero\ are available to further investigate the similarities and the possible connection between this source and the ``gamma-ray binaries'' \src, \ls, and \psr.
\begin{acknowledgements}
Based on observations with the NASA/UK/ASI Swift mission, obtained through the High Energy Astrophysics Science Archive Research Center Online Service, provided by the NASA/Goddard Space Flight Center. We thank the Swift team for making these observations possible, in particular the duty scientists and science planners.
The Italian authors acknowledge the support of the Italian Space Agency (contract ASI/INAF I/023/05/0) and the Italian Ministry for University and Research (grant PRIN 2005 02 5417). 
A.D.L. acknowledges an Italian Space Agency fellowship,
M.A.M. a ``G. Petrocchi'' fellowship of the Osio Sotto (BG) city council.
We thank Pat Romano for her help with the Swift/XRT data. We also thank Lara Sidoli, Ada Paizis, Andrea Tiengo, and Stefano Vercellone for useful discussions.
\end{acknowledgements}
\bibliographystyle{aa}
\bibliography{biblio}
\end{document}